\documentclass[jjap3,reprint,twocolumn,amsmath,amssymb]{revtex4-1}
\usepackage{stmaryrd}
\usepackage{times}
\usepackage{graphicx}
\usepackage{dcolumn}
\usepackage{bm}
\usepackage{float}

\begin{document}

\title{Separation of inverse spin Hall effect and anomalous Nernst effect in ferromagnetic metals}

\author{Hao Wu}
\affiliation{Beijing National Laboratory for Condensed Matter Physics, Institute of Physics, University of Chinese Academy of Sciences, Chinese Academy of Sciences, Beijing 100190, China}

\author{Xiao Wang}
\affiliation{Beijing National Laboratory for Condensed Matter Physics, Institute of Physics, University of Chinese Academy of Sciences, Chinese Academy of Sciences, Beijing 100190, China}

\author{Li Huang}
\affiliation{Beijing National Laboratory for Condensed Matter Physics, Institute of Physics, University of Chinese Academy of Sciences, Chinese Academy of Sciences, Beijing 100190, China}

\author{Jianying Qin}
\affiliation{Beijing National Laboratory for Condensed Matter Physics, Institute of Physics, University of Chinese Academy of Sciences, Chinese Academy of Sciences, Beijing 100190, China}

\author{Chi Fang}
\affiliation{Beijing National Laboratory for Condensed Matter Physics, Institute of Physics, University of Chinese Academy of Sciences, Chinese Academy of Sciences, Beijing 100190, China}

\author{Xuan Zhang}
\affiliation{Beijing National Laboratory for Condensed Matter Physics, Institute of Physics, University of Chinese Academy of Sciences, Chinese Academy of Sciences, Beijing 100190, China}

\author{Caihua Wan}
\affiliation{Beijing National Laboratory for Condensed Matter Physics, Institute of Physics, University of Chinese Academy of Sciences, Chinese Academy of Sciences, Beijing 100190, China}

\author{Xiufeng Han}
\email[Email: ]{xfhan@iphy.ac.cn}
\affiliation{Beijing National Laboratory for Condensed Matter Physics, Institute of Physics, University of Chinese Academy of Sciences, Chinese Academy of Sciences, Beijing 100190, China}

\begin{abstract}

Inverse spin Hall effect (ISHE) in ferromagnetic metals (FM) can also be used to detect the spin current generated by longitudinal spin Seebeck effect in a ferromagnetic insulator YIG. However, anomalous Nernst effect (ANE) in FM itself always mixes in the thermal voltage. In this work, the exchange bias structure (NiFe/IrMn) is employed to separate these two effects. The exchange bias structure provides a shift field to NiFe, which can separate the magnetization of NiFe from that of YIG in \emph{M}-\emph{H} loops. As a result, the ISHE related to magnetization of YIG and the ANE related to the magnetization of NiFe can be separated as well. By comparison with Pt, a relative spin Hall angle of NiFe (0.87) is obtained, which results from the partially filled 3\emph{d} orbits and the ferromagnetic order. This work puts forward a practical method to use the ISHE in ferromagnetic metals towards future spintronic applications.

\end{abstract}


\maketitle


Spin caloritronics focuses on coupling heat, spin and charge in magnetic materials. \cite{bauer2012spin} Spin Seebeck effect (SSE) origins from the excitation of spin wave in magnetic materials by a temperature gradient, which can pump a spin current into a contact metal. \cite{uchida2008observation,uchida2010spin,uchida2010observation,xiao2010theory} In the past several years, SSE has been achieved in magnetic metals\cite{uchida2008observation}, semiconductors\cite{jaworski2010observation} and insulators. \cite{uchida2010spin,uchida2010observation} Especially SSE in magnetic insulators draws many attentions since a pure spin current without any charge flow is one of the most desirable properties for devices with dramatically reduced power consumption. Transverse and longitudinal spin Seebeck effect are divided by different experimental configurations, while the detected spin current can be perpendicular or parallel to the temperature gradient. Especially longitudinal spin Seebeck effect (LSSE) in ferromagnetic insulators is widely used to pump a spin current into the neighboring materials.

Inverse spin Hall effect (ISHE) can convert the spin current $\textbf{\emph{J}}_{\mbox{\footnotesize s}}$ into the charge current $\textbf{\emph{J}}_{\mbox{\footnotesize e}}$, which can be detected by a voltage signal: $\textbf{\emph{E}}_{\mbox{\tiny ISHE}} =(\theta_{\mbox{\tiny SH}}\rho)\textbf{\emph{J}}_{\mbox{\footnotesize s}}\times\bm\sigma$, where $\textbf{\emph{E}}_{\mbox{\tiny ISHE}}$ is the ISHE electric field, $\theta_{\mbox{\tiny SH}}$ is the spin Hall angle, $\rho$ is the resistivity, and $\bm\sigma$ is the unit vector of the spin. \cite{valenzuela2006direct,kimura2007room,karplus1954hall} It is generally believed that the magnitude of spin Hall angle $\theta_{\mbox{\tiny SH}}$ depends on the strength of spin orbit coupling (SOC), and the strength of SOC is proportional to $Z^4$, while $Z$ is the atomic number, so heavy metals (HM) with large $Z$ have a relative large spin Hall angle. \cite{wang2014scaling}

Similar to spin Hall effect (SHE) in non-magnetic metals, \cite{hirsch1999spin,kato2004observation,mosendz2010quantifying} anomalous Hall effect (AHE) in ferromagnetic metals (FM) comes from the spin dependent scattering of the charge current. \cite{nagaosa2010anomalous} Due to the spin polarization of the charge current in FM, the spin accumulation accompanied with a charge accumulation can be generated in the transverse direction. When a pure spin current is injected to FM, as the inverse effect of AHE, ISHE in FM provides another potential application in detecting the spin current by charge signals.

Recent works draw attention on using the ISHE in FM to detect the spin current generated by LSSE in a ferromagnetic insulator Y$_3$Fe$_5$O$_{12}$ (YIG).\cite{miao2013inverse,wu2014unambiguous,tian2015separation,seki2015observation} However, the temperature gradient will also introduce additional anomalous Nernst effect (ANE) in FM: $\textbf{\emph{E}}_{\mbox{\tiny ANE}}\propto\nabla\textbf{\emph{T}}_{\mbox{\footnotesize {z}}}\times\textbf{\emph{M}}$, where $\textbf{\emph{E}}_{\mbox{\tiny ANE}}$ is the ANE electric field, $\nabla\textbf{\emph{T}}_{\mbox{\footnotesize{z}}}$ is the temperature gradient along the thickness direction, and $\textbf{\emph{M}}$ is the magnetic moment of FM.\cite{slachter2011anomalous,mizuguchi2012anomalous} Therefore, the separation of ANE and ISHE in FM is in great demand. Several works have used two magnetic materials with different coercivity to separate the ISHE (related to the magnetization of YIG) and ANE (related to the magnetization of the ferromagnetic detector).\cite{wu2014unambiguous,tian2015separation} However, ISHE and ANE are still mixed with each other, which prevents us to directly detect the spin current by FM.

\begin{figure*}
\includegraphics[width=110mm]{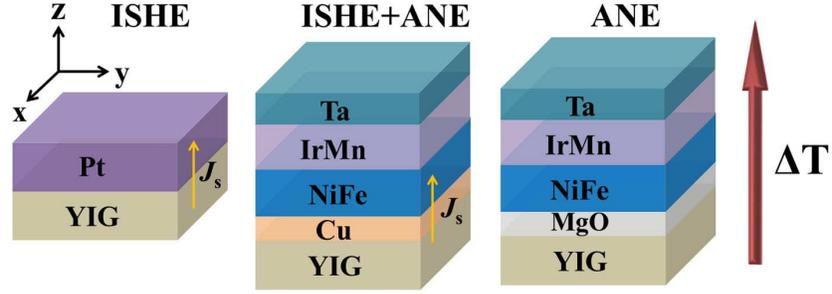}
\caption{\label{FIG1} Under the longitudinal temperature gradient, only the inverse spin Hall effect (ISHE) exists in YIG/Pt sample, while both ISHE and the anomalous Nernst effect (ANE) exist in YIG/Cu/NiFe/IrMn/Ta sample. Once an insulating layer MgO is inserted between NiFe and YIG, the spin current will be blocked, so ISHE vanishes while ANE still exists.}
\end{figure*}

In this work, we designed the exchange bias structure (NiFe/IrMn) to detect the spin current generated by LSSE in YIG. A Cu layer with a negligible spin Hall angle is inserted between NiFe and YIG to reduce the magnetic coupling between YIG and NiFe, and the spin current can also pass without too much loss at the same time. The exchange bias structure provides a bias field for NiFe, which can separate the magnetization switching process of NiFe from that of YIG in $\emph{M}$-$\emph{H}$ loops. \cite{koon1997calculations,berkowitz1999exchange} As a result, the ISHE related to magnetization of YIG and the ANE related to the magnetization of NiFe can be separated as well. More importantly, we can even observe the only ISHE contribution in a field range which is smaller than the exchange bias field that only the magnetization of YIG switches, while the magnetization of NiFe is fixed.

The exchange bias structure Cu(5)/NiFe(5)/IrMn(12)/Ta(5) (thickness in nanometers) was fabricated on polished 3.5 $\mu$m YIG films on GGG substrates by a magnetron sputtering system. In order to introduce the exchange bias effect in FM/AFM, an in-plane magnetic field was applied during the deposition process. The spin Seebeck voltage was measure by a nanovoltmeter (Keithley 2182A) in a Seebeck measurement system with a Helmholtz coil, and the longitudinal temperature difference along the thickness direction was measured between the bottom of the GGG substrate and the top of the film. All data was performed at room temperature.

\begin{figure*}
\includegraphics[width=120mm]{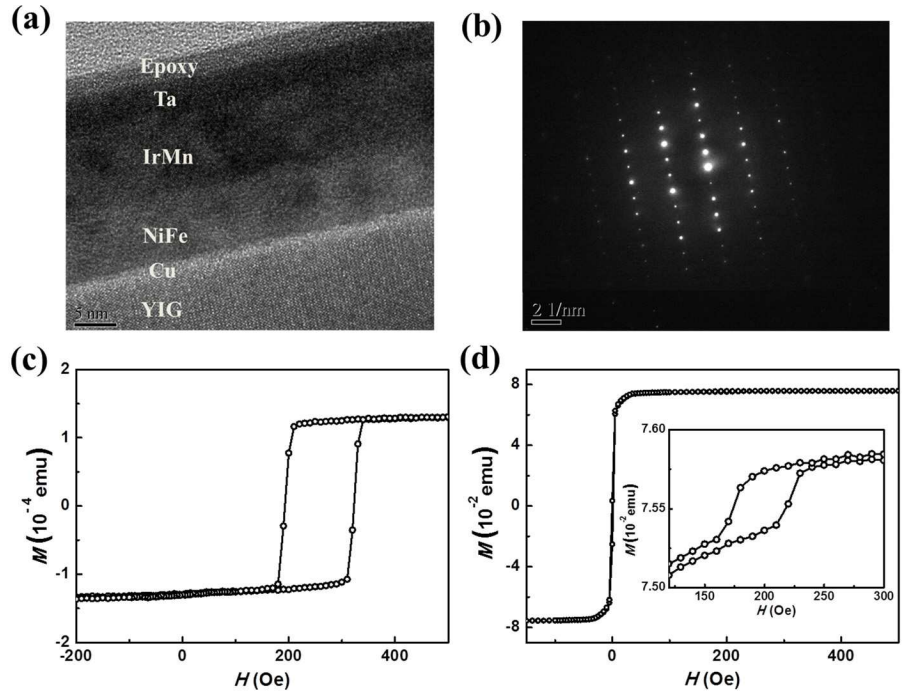}
\caption{\label{FIG2} High resolution transmission electron microscopy (HRTEM) results of the YIG/Cu/NiFe/IrMn/Ta sample (a) and selected area electron diffraction (SAED) patterns of the YIG region (b). \emph{M}-\emph{H} loops measured in Si-SiO$_{2}$/Cu/NiFe/IrMn/Ta (c) and YIG/Cu/NiFe/IrMn/Ta (d), and the magnetic field is applied along the direction of exchange bias field.}
\end{figure*}

Fig. 1(a) shows the schematic diagram of the measurement method and the physical process. In longitudinal spin Seebeck measurement, the temperature gradient ($\nabla\textbf{\emph{T}}$) is applied along the out-of-plane $z$ direction, and the magnetic field is scanned along $x$ direction (also the direction of the exchange bias field). According to $\textbf{\emph{E}}_{\mbox{\tiny ISHE}} =(\theta_{\mbox{\tiny SH}}\rho)\textbf{\emph{J}}_{\mbox{\footnotesize s}}\times\bm\sigma$, the thermal voltage should be measured along $y$ direction. Firstly, as previous works, we use a Pt layer which has a relative large spin Hall angle about 0.1 \cite{jiao2013spin,zhang2013determination,obstbaum2014inverse} to measure the pure spin Seebeck induced ISHE signal. Then, we changed the Pt with the exchange bias structure Cu/NiFe/IrMn/Ta. Apart from the ISHE signal, ANE from NiFe itself will also contribute to the thermal voltage. Once we inserted an insulating layer MgO to block the spin current injected from YIG to NiFe, ISHE signal should vanish where only ANE from NiFe exists.

\begin{figure}
\includegraphics[width=60mm]{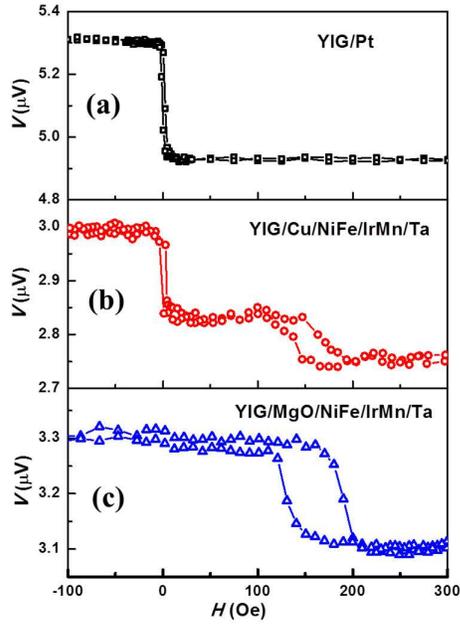}
\caption{\label{FIG3} (a)-(c) shows the spin dependent thermal voltage measurement of YIG/Pt, YIG/Cu/NiFe/IrMn/Ta, and YIG/MgO/NiFe/IrMn/Ta samples, where the magnetic field is applied along $x$ direction (also the direction of the exchange bias field).}
\end{figure}

Fig. 2(a) shows the cross-sectional transmission electron microscopy (HRTEM) results of the YIG/Cu/NiFe/IrMn/Ta multilayers, and the interface between Cu and YIG is very flat and clear. The selected area electron diffraction (SAED) pattern is shown in Fig. 2(b), demonstrating that the epitaxial direction of YIG film crystal is along the (111) direction and the lattice parameter is 12.4 {\AA}. A Si-SiO$_2$/Cu/NiFe/IrMn/Ta reference sample is used to check the exchange bias effect, where a 220 Oe exchange bias field is obtained from the $\emph{M}$-$\emph{H}$ loop (along $x$ direction) [Fig. 2(c)]. Then we measured the magnetic properties of YIG/Cu/NiFe/IrMn/Ta sample [Fig. 2(d)], and the saturation magnetization $\emph{M}_{\mbox{\footnotesize s}}$ of YIG is 120 emu/cc and the saturation field of YIG is less than 10 Oe. From the zoom-in figure in Fig. 2(d), we can see the magnetization switching range of NiFe/IrMn exchange bias structure is from 150 Oe to 250 Oe, which is far from the range of YIG.

Then we measured the magnetic field dependence of the thermal voltage, and the longitudinal temperature difference $\Delta \emph{T}$ keeps 13 K during the measurement in Fig. 3(a)-(c). Firstly, as conventional LSSE measurement, a heavy metal Pt is used to measure the spin current generated by SSE in YIG, and a 0.4 $\mu$V ISHE voltage which is related to the magnetization of YIG is observed, as shown in Fig. 3(a). While in YIG/Cu/NiFe/IrMn/Ta structure, due to the exchange bias effect, ISHE (related to the magnetization of YIG) and ANE (related to the magnetization of NiFe) exist in different field ranges. The ISHE signal is around 0 Oe and the ANE signal is around 150 Oe, and the exchange bias field here is a little smaller than that from the \emph{M}-\emph{H} loop because the exchange bias decreases with the increased temperature, as seen in Fig. 3(b). It is worth noting that the polarity of ISHE and ANE in NiFe is the same, which is contrary to the result for CoFeB. \cite{wu2014unambiguous} In order to prove that the ISHE voltage related to the magnetization of YIG indeed comes from the spin current injection from YIG to NiFe, we insert an insulating layer MgO to block this spin current. As expected, the ISHE signal disappears while the ANE still exists, as seen in Fig. 3(c). For SSE in heavy metal/ferromagnet structures, the mixing of magnetic proximity effect \cite{huang2012transport,lu2013pt} (MPE) and ANE has a debate for a long time, which means that magnetized Pt shows some ferromagnetic properties in transport measurement such as ANE and AHE. In our work, we directly use a FM to detect the spin current generated by SSE, and our results show that although both ISHE and ANE take place in the thermal voltage, however, by using the exchange bias effect, ISHE and ANE can be separated in different field ranges. These results demonstrate that SSE and ANE share different physical origins, and ANE is not the essential condition of SSE.

\begin{figure}
\includegraphics[width=60mm]{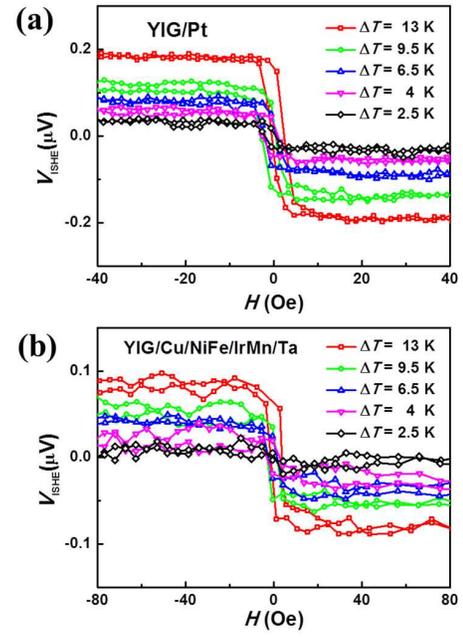}
\caption{\label{FIG4} (a) and (b) show the spin dependent thermal voltage measurement in YIG/Pt, YIG/Cu/NiFe/IrMn/Ta samples respectively under the varied longitudinal temperature difference from 2.5 K to 13 K, and the offset voltage has been removed to obtain the field dependent ISHE contribution. The magnetic field is applied along $x$ direction (also the direction of the exchange bias field), and the field range is smaller than the exchange bias field.}
\end{figure}

Then, we changed the temperature differences $\Delta \emph{T}$ from 2.5 K to 13K, and the field dependent ISHE voltage for different $\Delta \emph{T}$ is shown in Fig. 4(a) (YIG/Pt) and Fig. 4(b) (YIG/Cu/NiFe/IrMn/Ta). The ISHE voltages gradually increase with increasing the temperature gradient in both samples, which are in accordance with the spin Seebeck mechanism. Under a $\pm$80 Oe field range which is smaller than the exchange bias field, only the pure ISHE signals without ANE shows that the comparable utility of FM (NiFe) with conventional heavy metals (Pt) in detecting the spin current. Because in this case only the magnetization of YIG reverses, while the magnetization of NiFe keeps fixed. And the optimization of FM with large spin Hall angle will be an essential step towards future applications.

In order to compare the spin Hall angle in Pt and NiFe, the ISHE voltage is normalized by the resistance of the detecting electrode $\emph{R}$, and the $\emph{V}_{\mbox{\tiny ISHE}}/ \emph{R}$-$\Delta \emph{T}$ curve is fitted by the linear shape, as seen in Fig. 5. $\emph{V}_{\mbox{\tiny ISHE}}/ \emph{R}=\beta \theta_{\mbox{\tiny SH}}\Delta \emph{T}$, where $\beta$ represents the efficiency from thermal current to the detected spin current.
If we assume the same $\beta$ in YIG/Pt and YIG/Cu/NiFe/IrMn/Ta samples, we can calculate the relative spin Hall angle of NiFe: $\theta_{\mbox{\tiny SH}}$(NiFe)/$\theta_{\mbox{\tiny SH}}$(Pt)$\approx$0.87, which is close to our previous result (0.98) by transverse SSE measurement. \cite{wu2015observation}

\begin{figure}
\includegraphics[width=60mm]{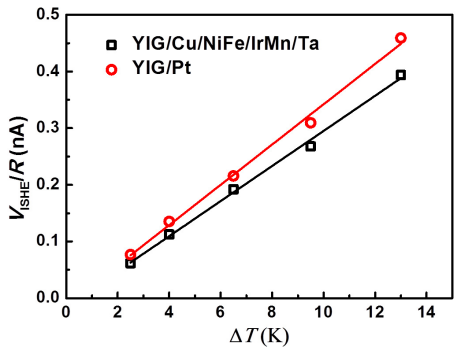}
\caption{\label{FIG5} The $\emph{V}_{\mbox{\tiny ISHE}}/ \emph{R}-\Delta \emph{T}$ curves and the linear fitting curves measured in YIG/Pt and YIG/Cu/NiFe/IrMn/Ta samples, and the ISHE voltage has been normalized by the resistance of the detecting electrode.}
\end{figure}

In conventional understanding of SOC, the strength of SOC follows a $Z^4$ dependence. While NiFe is composed of light atoms, so SOC in NiFe should be small in this mechanism. However, previous works have also shown that SOC not only depends on the atomic number \emph{Z} but also depends on the filling of \emph{d}-orbit, both Ni and Fe have partially filled 3\emph{d} orbits, so SOC from the \emph{d}-orbit filling could take an important role in NiFe. \cite{du2014systematic,morota2011indication} Moreover, ferromagnetic order induced intrinsic spin dependent scattering which is solely determined by the electronic band structure can also contribute to the ISHE in FM, because the ISHE in FM is independent of its magnetization. \cite{tian2016manipulation}

In conclusion, we have designed the exchange bias structure (NiFe/IrMn) to separate the ISHE and ANE in FM. As expected, the ISHE related to magnetization of YIG and the ANE related to the magnetization of NiFe can be separated in different ranges of magnetic field. By linear fitting the $\emph{V}_{\mbox{\tiny ISHE}}/ \emph{R}$-$\Delta \emph{T}$ curves of NiFe and Pt, we calculated the relative spin Hall angle $\theta_{\mbox{\tiny SH}}$(NiFe)/$\theta_{\mbox{\tiny SH}}$(Pt)=0.87, and the partial filling of 3\emph{d} orbits and the ferromagnetic order play important roles in this large spin Hall angle of NiFe. This work demonstrates that ferromagnetic metals can also be used to detect the spin current in spintronics devices.

\begin{acknowledgments}
This work was supported by the 863 Plan Project of Ministry of Science and Technology (MOST) [Grant No. 2014AA032904], the National Key Research and Development Program of China [Grant No. 2016YFA0300802], the National Natural Science Foundation of China (NSFC) [Grant Nos. 11434014, 11404382], and the Strategic Priority Research Program (B) of the Chinese Academy of Sciences (CAS) [Grant No. XDB07030200].
\end{acknowledgments}

%


\begin{thebibliography}{31}%
\makeatletter
\providecommand \@ifxundefined [1]{%
 \@ifx{#1\undefined}
}%
\providecommand \@ifnum [1]{%
 \ifnum #1\expandafter \@firstoftwo
 \else \expandafter \@secondoftwo
 \fi
}%
\providecommand \@ifx [1]{%
 \ifx #1\expandafter \@firstoftwo
 \else \expandafter \@secondoftwo
 \fi
}%
\providecommand \natexlab [1]{#1}%
\providecommand \enquote  [1]{``#1''}%
\providecommand \bibnamefont  [1]{#1}%
\providecommand \bibfnamefont [1]{#1}%
\providecommand \citenamefont [1]{#1}%
\providecommand \href@noop [0]{\@secondoftwo}%
\providecommand \href [0]{\begingroup \@sanitize@url \@href}%
\providecommand \@href[1]{\@@startlink{#1}\@@href}%
\providecommand \@@href[1]{\endgroup#1\@@endlink}%
\providecommand \@sanitize@url [0]{\catcode `\\12\catcode `\$12\catcode
  `\&12\catcode `\#12\catcode `\^12\catcode `\_12\catcode `\%12\relax}%
\providecommand \@@startlink[1]{}%
\providecommand \@@endlink[0]{}%
\providecommand \url  [0]{\begingroup\@sanitize@url \@url }%
\providecommand \@url [1]{\endgroup\@href {#1}{\urlprefix }}%
\providecommand \urlprefix  [0]{URL }%
\providecommand \Eprint [0]{\href }%
\providecommand \doibase [0]{http://dx.doi.org/}%
\providecommand \selectlanguage [0]{\@gobble}%
\providecommand \bibinfo  [0]{\@secondoftwo}%
\providecommand \bibfield  [0]{\@secondoftwo}%
\providecommand \translation [1]{[#1]}%
\providecommand \BibitemOpen [0]{}%
\providecommand \bibitemStop [0]{}%
\providecommand \bibitemNoStop [0]{.\EOS\space}%
\providecommand \EOS [0]{\spacefactor3000\relax}%
\providecommand \BibitemShut  [1]{\csname bibitem#1\endcsname}%
\let\auto@bib@innerbib\@empty
\bibitem [{\citenamefont {Bauer}\ \emph {et~al.}(2012)\citenamefont {Bauer},
  \citenamefont {Saitoh},\ and\ \citenamefont {van Wees}}]{bauer2012spin}%
  \BibitemOpen
  \bibfield  {author} {\bibinfo {author} {\bibfnamefont {G.~E.~W.}\
  \bibnamefont {Bauer}}, \bibinfo {author} {\bibfnamefont {E.}~\bibnamefont
  {Saitoh}}, \ and\ \bibinfo {author} {\bibfnamefont {B.~J.}\ \bibnamefont {van
  Wees}},\ }\href@noop {} {\bibfield  {journal} {\bibinfo  {journal} {Nat.
  Mater.}\ }\textbf {\bibinfo {volume} {11}},\ \bibinfo {pages} {391} (\bibinfo
  {year} {2012})}\BibitemShut {NoStop}%
\bibitem [{\citenamefont {Uchida}\ \emph {et~al.}(2008)\citenamefont {Uchida},
  \citenamefont {Takahashi}, \citenamefont {Harii}, \citenamefont {Ieda},
  \citenamefont {Koshibae}, \citenamefont {Ando}, \citenamefont {Maekawa},\
  and\ \citenamefont {Saitoh}}]{uchida2008observation}%
  \BibitemOpen
  \bibfield  {author} {\bibinfo {author} {\bibfnamefont {K.}~\bibnamefont
  {Uchida}}, \bibinfo {author} {\bibfnamefont {S.}~\bibnamefont {Takahashi}},
  \bibinfo {author} {\bibfnamefont {K.}~\bibnamefont {Harii}}, \bibinfo
  {author} {\bibfnamefont {J.}~\bibnamefont {Ieda}}, \bibinfo {author}
  {\bibfnamefont {W.}~\bibnamefont {Koshibae}}, \bibinfo {author}
  {\bibfnamefont {K.}~\bibnamefont {Ando}}, \bibinfo {author} {\bibfnamefont
  {S.}~\bibnamefont {Maekawa}}, \ and\ \bibinfo {author} {\bibfnamefont
  {E.}~\bibnamefont {Saitoh}},\ }\href@noop {} {\bibfield  {journal} {\bibinfo
  {journal} {Nature}\ }\textbf {\bibinfo {volume} {455}},\ \bibinfo {pages}
  {778} (\bibinfo {year} {2008})}\BibitemShut {NoStop}%
\bibitem [{\citenamefont {Uchida}\ \emph
  {et~al.}(2010{\natexlab{a}})\citenamefont {Uchida}, \citenamefont {Xiao},
  \citenamefont {Adachi}, \citenamefont {Ohe}, \citenamefont {Takahashi},
  \citenamefont {Ieda}, \citenamefont {Ota}, \citenamefont {Kajiwara},
  \citenamefont {Umezawa}, \citenamefont {Kawai}, \citenamefont {Bauer},
  \citenamefont {Maekawa},\ and\ \citenamefont {Saitoh}}]{uchida2010spin}%
  \BibitemOpen
  \bibfield  {author} {\bibinfo {author} {\bibfnamefont {K.}~\bibnamefont
  {Uchida}}, \bibinfo {author} {\bibfnamefont {J.}~\bibnamefont {Xiao}},
  \bibinfo {author} {\bibfnamefont {H.}~\bibnamefont {Adachi}}, \bibinfo
  {author} {\bibfnamefont {J.}~\bibnamefont {Ohe}}, \bibinfo {author}
  {\bibfnamefont {S.}~\bibnamefont {Takahashi}}, \bibinfo {author}
  {\bibfnamefont {J.}~\bibnamefont {Ieda}}, \bibinfo {author} {\bibfnamefont
  {T.}~\bibnamefont {Ota}}, \bibinfo {author} {\bibfnamefont {Y.}~\bibnamefont
  {Kajiwara}}, \bibinfo {author} {\bibfnamefont {H.}~\bibnamefont {Umezawa}},
  \bibinfo {author} {\bibfnamefont {H.}~\bibnamefont {Kawai}}, \bibinfo
  {author} {\bibfnamefont {G.~E.~W.}\ \bibnamefont {Bauer}}, \bibinfo {author}
  {\bibfnamefont {S.}~\bibnamefont {Maekawa}}, \ and\ \bibinfo {author}
  {\bibfnamefont {E.}~\bibnamefont {Saitoh}},\ }\href@noop {} {\bibfield
  {journal} {\bibinfo  {journal} {Nat. Mater.}\ }\textbf {\bibinfo {volume}
  {9}},\ \bibinfo {pages} {894} (\bibinfo {year}
  {2010}{\natexlab{a}})}\BibitemShut {NoStop}%
\bibitem [{\citenamefont {Uchida}\ \emph
  {et~al.}(2010{\natexlab{b}})\citenamefont {Uchida}, \citenamefont {Adachi},
  \citenamefont {Ota}, \citenamefont {Nakayama}, \citenamefont {Maekawa},\ and\
  \citenamefont {Saitoh}}]{uchida2010observation}%
  \BibitemOpen
  \bibfield  {author} {\bibinfo {author} {\bibfnamefont {K.}~\bibnamefont
  {Uchida}}, \bibinfo {author} {\bibfnamefont {H.}~\bibnamefont {Adachi}},
  \bibinfo {author} {\bibfnamefont {T.}~\bibnamefont {Ota}}, \bibinfo {author}
  {\bibfnamefont {H.}~\bibnamefont {Nakayama}}, \bibinfo {author}
  {\bibfnamefont {S.}~\bibnamefont {Maekawa}}, \ and\ \bibinfo {author}
  {\bibfnamefont {E.}~\bibnamefont {Saitoh}},\ }\href@noop {} {\bibfield
  {journal} {\bibinfo  {journal} {Appl. Phys. Lett.}\ }\textbf {\bibinfo
  {volume} {97}},\ \bibinfo {pages} {172505} (\bibinfo {year}
  {2010}{\natexlab{b}})}\BibitemShut {NoStop}%
\bibitem [{\citenamefont {Xiao}\ \emph {et~al.}(2010)\citenamefont {Xiao},
  \citenamefont {Bauer}, \citenamefont {Uchida}, \citenamefont {Saitoh},\ and\
  \citenamefont {Maekawa}}]{xiao2010theory}%
  \BibitemOpen
  \bibfield  {author} {\bibinfo {author} {\bibfnamefont {J.}~\bibnamefont
  {Xiao}}, \bibinfo {author} {\bibfnamefont {G.~E.~W.}\ \bibnamefont {Bauer}},
  \bibinfo {author} {\bibfnamefont {K.}~\bibnamefont {Uchida}}, \bibinfo
  {author} {\bibfnamefont {E.}~\bibnamefont {Saitoh}}, \ and\ \bibinfo {author}
  {\bibfnamefont {S.}~\bibnamefont {Maekawa}},\ }\href@noop {} {\bibfield
  {journal} {\bibinfo  {journal} {Phys. Rev. B}\ }\textbf {\bibinfo {volume}
  {81}},\ \bibinfo {pages} {214418} (\bibinfo {year} {2010})}\BibitemShut
  {NoStop}%
\bibitem [{\citenamefont {Jaworski}\ \emph {et~al.}(2010)\citenamefont
  {Jaworski}, \citenamefont {Yang}, \citenamefont {Mack}, \citenamefont
  {Awschalom}, \citenamefont {Heremans},\ and\ \citenamefont
  {Myers}}]{jaworski2010observation}%
  \BibitemOpen
  \bibfield  {author} {\bibinfo {author} {\bibfnamefont {C.}~\bibnamefont
  {Jaworski}}, \bibinfo {author} {\bibfnamefont {J.}~\bibnamefont {Yang}},
  \bibinfo {author} {\bibfnamefont {S.}~\bibnamefont {Mack}}, \bibinfo {author}
  {\bibfnamefont {D.}~\bibnamefont {Awschalom}}, \bibinfo {author}
  {\bibfnamefont {J.}~\bibnamefont {Heremans}}, \ and\ \bibinfo {author}
  {\bibfnamefont {R.}~\bibnamefont {Myers}},\ }\href@noop {} {\bibfield
  {journal} {\bibinfo  {journal} {Nat. Mater.}\ }\textbf {\bibinfo {volume}
  {9}},\ \bibinfo {pages} {898} (\bibinfo {year} {2010})}\BibitemShut {NoStop}%
\bibitem [{\citenamefont {Valenzuela}\ and\ \citenamefont
  {Tinkham}(2006)}]{valenzuela2006direct}%
  \BibitemOpen
  \bibfield  {author} {\bibinfo {author} {\bibfnamefont {S.~O.}\ \bibnamefont
  {Valenzuela}}\ and\ \bibinfo {author} {\bibfnamefont {M.}~\bibnamefont
  {Tinkham}},\ }\href@noop {} {\bibfield  {journal} {\bibinfo  {journal}
  {Nature}\ }\textbf {\bibinfo {volume} {442}},\ \bibinfo {pages} {176}
  (\bibinfo {year} {2006})}\BibitemShut {NoStop}%
\bibitem [{\citenamefont {Kimura}\ \emph {et~al.}(2007)\citenamefont {Kimura},
  \citenamefont {Otani}, \citenamefont {Sato}, \citenamefont {Takahashi},\ and\
  \citenamefont {Maekawa}}]{kimura2007room}%
  \BibitemOpen
  \bibfield  {author} {\bibinfo {author} {\bibfnamefont {T.}~\bibnamefont
  {Kimura}}, \bibinfo {author} {\bibfnamefont {Y.}~\bibnamefont {Otani}},
  \bibinfo {author} {\bibfnamefont {T.}~\bibnamefont {Sato}}, \bibinfo {author}
  {\bibfnamefont {S.}~\bibnamefont {Takahashi}}, \ and\ \bibinfo {author}
  {\bibfnamefont {S.}~\bibnamefont {Maekawa}},\ }\href@noop {} {\bibfield
  {journal} {\bibinfo  {journal} {Phys. Rev. Lett.}\ }\textbf {\bibinfo
  {volume} {98}},\ \bibinfo {pages} {156601} (\bibinfo {year}
  {2007})}\BibitemShut {NoStop}%
\bibitem [{\citenamefont {Karplus}\ and\ \citenamefont
  {Luttinger}(1954)}]{karplus1954hall}%
  \BibitemOpen
  \bibfield  {author} {\bibinfo {author} {\bibfnamefont {R.}~\bibnamefont
  {Karplus}}\ and\ \bibinfo {author} {\bibfnamefont {J.}~\bibnamefont
  {Luttinger}},\ }\href@noop {} {\bibfield  {journal} {\bibinfo  {journal}
  {Phys. Rev.}\ }\textbf {\bibinfo {volume} {95}},\ \bibinfo {pages} {1154}
  (\bibinfo {year} {1954})}\BibitemShut {NoStop}%
\bibitem [{\citenamefont {Wang}\ \emph {et~al.}(2014)\citenamefont {Wang},
  \citenamefont {Du}, \citenamefont {Pu}, \citenamefont {Adur}, \citenamefont
  {Hammel},\ and\ \citenamefont {Yang}}]{wang2014scaling}%
  \BibitemOpen
  \bibfield  {author} {\bibinfo {author} {\bibfnamefont {H.~L.}\ \bibnamefont
  {Wang}}, \bibinfo {author} {\bibfnamefont {C.~H.}\ \bibnamefont {Du}},
  \bibinfo {author} {\bibfnamefont {Y.}~\bibnamefont {Pu}}, \bibinfo {author}
  {\bibfnamefont {R.}~\bibnamefont {Adur}}, \bibinfo {author} {\bibfnamefont
  {P.~C.}\ \bibnamefont {Hammel}}, \ and\ \bibinfo {author} {\bibfnamefont
  {F.~Y.}\ \bibnamefont {Yang}},\ }\href@noop {} {\bibfield  {journal}
  {\bibinfo  {journal} {Phys. Rev. Lett.}\ }\textbf {\bibinfo {volume} {112}},\
  \bibinfo {pages} {197201} (\bibinfo {year} {2014})}\BibitemShut {NoStop}%
\bibitem [{\citenamefont {Hirsch}(1999)}]{hirsch1999spin}%
  \BibitemOpen
  \bibfield  {author} {\bibinfo {author} {\bibfnamefont {J.~E.}\ \bibnamefont
  {Hirsch}},\ }\href@noop {} {\bibfield  {journal} {\bibinfo  {journal} {Phys.
  Rev. Lett.}\ }\textbf {\bibinfo {volume} {83}},\ \bibinfo {pages} {1834}
  (\bibinfo {year} {1999})}\BibitemShut {NoStop}%
\bibitem [{\citenamefont {Kato}\ \emph {et~al.}(2004)\citenamefont {Kato},
  \citenamefont {Myers}, \citenamefont {Gossard},\ and\ \citenamefont
  {Awschalom}}]{kato2004observation}%
  \BibitemOpen
  \bibfield  {author} {\bibinfo {author} {\bibfnamefont {Y.}~\bibnamefont
  {Kato}}, \bibinfo {author} {\bibfnamefont {R.}~\bibnamefont {Myers}},
  \bibinfo {author} {\bibfnamefont {A.}~\bibnamefont {Gossard}}, \ and\
  \bibinfo {author} {\bibfnamefont {D.}~\bibnamefont {Awschalom}},\ }\href@noop
  {} {\bibfield  {journal} {\bibinfo  {journal} {Science}\ }\textbf {\bibinfo
  {volume} {306}},\ \bibinfo {pages} {1910} (\bibinfo {year}
  {2004})}\BibitemShut {NoStop}%
\bibitem [{\citenamefont {Mosendz}\ \emph {et~al.}(2010)\citenamefont
  {Mosendz}, \citenamefont {Pearson}, \citenamefont {Fradin}, \citenamefont
  {Bauer}, \citenamefont {Bader},\ and\ \citenamefont
  {Hoffmann}}]{mosendz2010quantifying}%
  \BibitemOpen
  \bibfield  {author} {\bibinfo {author} {\bibfnamefont {O.}~\bibnamefont
  {Mosendz}}, \bibinfo {author} {\bibfnamefont {J.}~\bibnamefont {Pearson}},
  \bibinfo {author} {\bibfnamefont {F.}~\bibnamefont {Fradin}}, \bibinfo
  {author} {\bibfnamefont {G.~E.~W.}\ \bibnamefont {Bauer}}, \bibinfo {author}
  {\bibfnamefont {S.}~\bibnamefont {Bader}}, \ and\ \bibinfo {author}
  {\bibfnamefont {A.}~\bibnamefont {Hoffmann}},\ }\href@noop {} {\bibfield
  {journal} {\bibinfo  {journal} {Phys. Rev. Lett.}\ }\textbf {\bibinfo
  {volume} {104}},\ \bibinfo {pages} {046601} (\bibinfo {year}
  {2010})}\BibitemShut {NoStop}%
\bibitem [{\citenamefont {Nagaosa}\ \emph {et~al.}(2010)\citenamefont
  {Nagaosa}, \citenamefont {Sinova}, \citenamefont {Onoda}, \citenamefont
  {MacDonald},\ and\ \citenamefont {Ong}}]{nagaosa2010anomalous}%
  \BibitemOpen
  \bibfield  {author} {\bibinfo {author} {\bibfnamefont {N.}~\bibnamefont
  {Nagaosa}}, \bibinfo {author} {\bibfnamefont {J.}~\bibnamefont {Sinova}},
  \bibinfo {author} {\bibfnamefont {S.}~\bibnamefont {Onoda}}, \bibinfo
  {author} {\bibfnamefont {A.}~\bibnamefont {MacDonald}}, \ and\ \bibinfo
  {author} {\bibfnamefont {N.}~\bibnamefont {Ong}},\ }\href@noop {} {\bibfield
  {journal} {\bibinfo  {journal} {Rev. Mod. Phys.}\ }\textbf {\bibinfo {volume}
  {82}},\ \bibinfo {pages} {1539} (\bibinfo {year} {2010})}\BibitemShut
  {NoStop}%
\bibitem [{\citenamefont {Miao}\ \emph {et~al.}(2013)\citenamefont {Miao},
  \citenamefont {Huang}, \citenamefont {Qu},\ and\ \citenamefont
  {Chien}}]{miao2013inverse}%
  \BibitemOpen
  \bibfield  {author} {\bibinfo {author} {\bibfnamefont {B.~F.}\ \bibnamefont
  {Miao}}, \bibinfo {author} {\bibfnamefont {S.~Y.}\ \bibnamefont {Huang}},
  \bibinfo {author} {\bibfnamefont {D.}~\bibnamefont {Qu}}, \ and\ \bibinfo
  {author} {\bibfnamefont {C.~L.}\ \bibnamefont {Chien}},\ }\href@noop {}
  {\bibfield  {journal} {\bibinfo  {journal} {Phys. Rev. Lett.}\ }\textbf
  {\bibinfo {volume} {111}},\ \bibinfo {pages} {066602} (\bibinfo {year}
  {2013})}\BibitemShut {NoStop}%
\bibitem [{\citenamefont {Wu}\ \emph {et~al.}(2014)\citenamefont {Wu},
  \citenamefont {Hoffman}, \citenamefont {Pearson},\ and\ \citenamefont
  {Bhattacharya}}]{wu2014unambiguous}%
  \BibitemOpen
  \bibfield  {author} {\bibinfo {author} {\bibfnamefont {S.~M.}\ \bibnamefont
  {Wu}}, \bibinfo {author} {\bibfnamefont {J.}~\bibnamefont {Hoffman}},
  \bibinfo {author} {\bibfnamefont {J.~E.}\ \bibnamefont {Pearson}}, \ and\
  \bibinfo {author} {\bibfnamefont {A.}~\bibnamefont {Bhattacharya}},\
  }\href@noop {} {\bibfield  {journal} {\bibinfo  {journal} {Appl. Phys.
  Lett.}\ }\textbf {\bibinfo {volume} {105}},\ \bibinfo {pages} {092409}
  (\bibinfo {year} {2014})}\BibitemShut {NoStop}%
\bibitem [{\citenamefont {Tian}\ \emph {et~al.}(2015)\citenamefont {Tian},
  \citenamefont {Li}, \citenamefont {Qu}, \citenamefont {Jin},\ and\
  \citenamefont {Chien}}]{tian2015separation}%
  \BibitemOpen
  \bibfield  {author} {\bibinfo {author} {\bibfnamefont {D.}~\bibnamefont
  {Tian}}, \bibinfo {author} {\bibfnamefont {Y.~F.}\ \bibnamefont {Li}},
  \bibinfo {author} {\bibfnamefont {D.}~\bibnamefont {Qu}}, \bibinfo {author}
  {\bibfnamefont {X.~F.}\ \bibnamefont {Jin}}, \ and\ \bibinfo {author}
  {\bibfnamefont {C.~L.}\ \bibnamefont {Chien}},\ }\href@noop {} {\bibfield
  {journal} {\bibinfo  {journal} {Appl. Phys. Lett.}\ }\textbf {\bibinfo
  {volume} {106}},\ \bibinfo {pages} {212407} (\bibinfo {year}
  {2015})}\BibitemShut {NoStop}%
\bibitem [{\citenamefont {Seki}\ \emph {et~al.}(2015)\citenamefont {Seki},
  \citenamefont {Uchida}, \citenamefont {Kikkawa}, \citenamefont {Qiu},
  \citenamefont {Saitoh},\ and\ \citenamefont
  {Takanashi}}]{seki2015observation}%
  \BibitemOpen
  \bibfield  {author} {\bibinfo {author} {\bibfnamefont {T.}~\bibnamefont
  {Seki}}, \bibinfo {author} {\bibfnamefont {K.}~\bibnamefont {Uchida}},
  \bibinfo {author} {\bibfnamefont {T.}~\bibnamefont {Kikkawa}}, \bibinfo
  {author} {\bibfnamefont {Z.}~\bibnamefont {Qiu}}, \bibinfo {author}
  {\bibfnamefont {E.}~\bibnamefont {Saitoh}}, \ and\ \bibinfo {author}
  {\bibfnamefont {K.}~\bibnamefont {Takanashi}},\ }\href@noop {} {\bibfield
  {journal} {\bibinfo  {journal} {Appl. Phys. Lett.}\ }\textbf {\bibinfo
  {volume} {107}},\ \bibinfo {pages} {092401} (\bibinfo {year}
  {2015})}\BibitemShut {NoStop}%
\bibitem [{\citenamefont {Slachter}\ \emph {et~al.}(2011)\citenamefont
  {Slachter}, \citenamefont {Bakker},\ and\ \citenamefont {van
  Wees}}]{slachter2011anomalous}%
  \BibitemOpen
  \bibfield  {author} {\bibinfo {author} {\bibfnamefont {A.}~\bibnamefont
  {Slachter}}, \bibinfo {author} {\bibfnamefont {F.~L.}\ \bibnamefont
  {Bakker}}, \ and\ \bibinfo {author} {\bibfnamefont {B.~J.}\ \bibnamefont {van
  Wees}},\ }\href@noop {} {\bibfield  {journal} {\bibinfo  {journal} {Phys.
  Rev. B}\ }\textbf {\bibinfo {volume} {84}},\ \bibinfo {pages} {020412}
  (\bibinfo {year} {2011})}\BibitemShut {NoStop}%
\bibitem [{\citenamefont {Mizuguchi}\ \emph {et~al.}(2012)\citenamefont
  {Mizuguchi}, \citenamefont {Ohata}, \citenamefont {Uchida}, \citenamefont
  {Saitoh},\ and\ \citenamefont {Takanashi}}]{mizuguchi2012anomalous}%
  \BibitemOpen
  \bibfield  {author} {\bibinfo {author} {\bibfnamefont {M.}~\bibnamefont
  {Mizuguchi}}, \bibinfo {author} {\bibfnamefont {S.}~\bibnamefont {Ohata}},
  \bibinfo {author} {\bibfnamefont {K.}~\bibnamefont {Uchida}}, \bibinfo
  {author} {\bibfnamefont {E.}~\bibnamefont {Saitoh}}, \ and\ \bibinfo {author}
  {\bibfnamefont {K.}~\bibnamefont {Takanashi}},\ }\href@noop {} {\bibfield
  {journal} {\bibinfo  {journal} {Appl. Phys. Express}\ }\textbf {\bibinfo
  {volume} {5}},\ \bibinfo {pages} {093002} (\bibinfo {year}
  {2012})}\BibitemShut {NoStop}%
\bibitem [{\citenamefont {Koon}(1997)}]{koon1997calculations}%
  \BibitemOpen
  \bibfield  {author} {\bibinfo {author} {\bibfnamefont {N.}~\bibnamefont
  {Koon}},\ }\href@noop {} {\bibfield  {journal} {\bibinfo  {journal} {Phys.
  Rev. Lett.}\ }\textbf {\bibinfo {volume} {78}},\ \bibinfo {pages} {4865}
  (\bibinfo {year} {1997})}\BibitemShut {NoStop}%
\bibitem [{\citenamefont {Berkowitz}\ and\ \citenamefont
  {Takano}(1999)}]{berkowitz1999exchange}%
  \BibitemOpen
  \bibfield  {author} {\bibinfo {author} {\bibfnamefont {A.}~\bibnamefont
  {Berkowitz}}\ and\ \bibinfo {author} {\bibfnamefont {K.}~\bibnamefont
  {Takano}},\ }\href@noop {} {\bibfield  {journal} {\bibinfo  {journal} {J.
  Magn. Magn. Mater.}\ }\textbf {\bibinfo {volume} {200}},\ \bibinfo {pages}
  {552} (\bibinfo {year} {1999})}\BibitemShut {NoStop}%
\bibitem [{\citenamefont {Jiao}\ and\ \citenamefont
  {Bauer}(2013)}]{jiao2013spin}%
  \BibitemOpen
  \bibfield  {author} {\bibinfo {author} {\bibfnamefont {H.}~\bibnamefont
  {Jiao}}\ and\ \bibinfo {author} {\bibfnamefont {G.~E.~W.}\ \bibnamefont
  {Bauer}},\ }\href@noop {} {\bibfield  {journal} {\bibinfo  {journal} {Phys.
  Rev. Lett.}\ }\textbf {\bibinfo {volume} {110}},\ \bibinfo {pages} {217602}
  (\bibinfo {year} {2013})}\BibitemShut {NoStop}%
\bibitem [{\citenamefont {Zhang}\ \emph {et~al.}(2013)\citenamefont {Zhang},
  \citenamefont {Vlaminck}, \citenamefont {Pearson}, \citenamefont {Divan},
  \citenamefont {Bader},\ and\ \citenamefont
  {Hoffmann}}]{zhang2013determination}%
  \BibitemOpen
  \bibfield  {author} {\bibinfo {author} {\bibfnamefont {W.}~\bibnamefont
  {Zhang}}, \bibinfo {author} {\bibfnamefont {V.}~\bibnamefont {Vlaminck}},
  \bibinfo {author} {\bibfnamefont {J.~E.}\ \bibnamefont {Pearson}}, \bibinfo
  {author} {\bibfnamefont {R.}~\bibnamefont {Divan}}, \bibinfo {author}
  {\bibfnamefont {S.~D.}\ \bibnamefont {Bader}}, \ and\ \bibinfo {author}
  {\bibfnamefont {A.}~\bibnamefont {Hoffmann}},\ }\href@noop {} {\bibfield
  {journal} {\bibinfo  {journal} {Appl. Phys. Lett.}\ }\textbf {\bibinfo
  {volume} {103}},\ \bibinfo {pages} {242414} (\bibinfo {year}
  {2013})}\BibitemShut {NoStop}%
\bibitem [{\citenamefont {Obstbaum}\ \emph {et~al.}(2014)\citenamefont
  {Obstbaum}, \citenamefont {H{\"a}rtinger}, \citenamefont {Bauer},
  \citenamefont {Meier}, \citenamefont {Swientek}, \citenamefont {Back},\ and\
  \citenamefont {Woltersdorf}}]{obstbaum2014inverse}%
  \BibitemOpen
  \bibfield  {author} {\bibinfo {author} {\bibfnamefont {M.}~\bibnamefont
  {Obstbaum}}, \bibinfo {author} {\bibfnamefont {M.}~\bibnamefont
  {H{\"a}rtinger}}, \bibinfo {author} {\bibfnamefont {H.}~\bibnamefont
  {Bauer}}, \bibinfo {author} {\bibfnamefont {T.}~\bibnamefont {Meier}},
  \bibinfo {author} {\bibfnamefont {F.}~\bibnamefont {Swientek}}, \bibinfo
  {author} {\bibfnamefont {C.}~\bibnamefont {Back}}, \ and\ \bibinfo {author}
  {\bibfnamefont {G.}~\bibnamefont {Woltersdorf}},\ }\href@noop {} {\bibfield
  {journal} {\bibinfo  {journal} {Phys. Rev. B}\ }\textbf {\bibinfo {volume}
  {89}},\ \bibinfo {pages} {060407} (\bibinfo {year} {2014})}\BibitemShut
  {NoStop}%
\bibitem [{\citenamefont {Huang}\ \emph {et~al.}(2012)\citenamefont {Huang},
  \citenamefont {Fan}, \citenamefont {Qu}, \citenamefont {Chen}, \citenamefont
  {Wang}, \citenamefont {Wu}, \citenamefont {Chen}, \citenamefont {Xiao},\ and\
  \citenamefont {Chien}}]{huang2012transport}%
  \BibitemOpen
  \bibfield  {author} {\bibinfo {author} {\bibfnamefont {S.~Y.}\ \bibnamefont
  {Huang}}, \bibinfo {author} {\bibfnamefont {X.}~\bibnamefont {Fan}}, \bibinfo
  {author} {\bibfnamefont {D.}~\bibnamefont {Qu}}, \bibinfo {author}
  {\bibfnamefont {Y.~P.}\ \bibnamefont {Chen}}, \bibinfo {author}
  {\bibfnamefont {W.~G.}\ \bibnamefont {Wang}}, \bibinfo {author}
  {\bibfnamefont {J.}~\bibnamefont {Wu}}, \bibinfo {author} {\bibfnamefont
  {T.~Y.}\ \bibnamefont {Chen}}, \bibinfo {author} {\bibfnamefont {J.~Q.}\
  \bibnamefont {Xiao}}, \ and\ \bibinfo {author} {\bibfnamefont {C.~L.}\
  \bibnamefont {Chien}},\ }\href@noop {} {\bibfield  {journal} {\bibinfo
  {journal} {Phys. Rev. Lett.}\ }\textbf {\bibinfo {volume} {109}},\ \bibinfo
  {pages} {107204} (\bibinfo {year} {2012})}\BibitemShut {NoStop}%
\bibitem [{\citenamefont {Lu}\ \emph {et~al.}(2013)\citenamefont {Lu},
  \citenamefont {Choi}, \citenamefont {Ortega}, \citenamefont {Cheng},
  \citenamefont {Cai}, \citenamefont {Huang}, \citenamefont {Sun},\ and\
  \citenamefont {Chien}}]{lu2013pt}%
  \BibitemOpen
  \bibfield  {author} {\bibinfo {author} {\bibfnamefont {Y.~M.}\ \bibnamefont
  {Lu}}, \bibinfo {author} {\bibfnamefont {Y.}~\bibnamefont {Choi}}, \bibinfo
  {author} {\bibfnamefont {C.~M.}\ \bibnamefont {Ortega}}, \bibinfo {author}
  {\bibfnamefont {X.~M.}\ \bibnamefont {Cheng}}, \bibinfo {author}
  {\bibfnamefont {J.~W.}\ \bibnamefont {Cai}}, \bibinfo {author} {\bibfnamefont
  {S.~Y.}\ \bibnamefont {Huang}}, \bibinfo {author} {\bibfnamefont
  {L.}~\bibnamefont {Sun}}, \ and\ \bibinfo {author} {\bibfnamefont {C.~L.}\
  \bibnamefont {Chien}},\ }\href@noop {} {\bibfield  {journal} {\bibinfo
  {journal} {Phys. Rev. Lett.}\ }\textbf {\bibinfo {volume} {110}},\ \bibinfo
  {pages} {147207} (\bibinfo {year} {2013})}\BibitemShut {NoStop}%
\bibitem [{\citenamefont {Wu}\ \emph {et~al.}(2015)\citenamefont {Wu},
  \citenamefont {Wan}, \citenamefont {Yuan}, \citenamefont {Zhang},
  \citenamefont {Jiang}, \citenamefont {Zhang}, \citenamefont {Wen},\ and\
  \citenamefont {Han}}]{wu2015observation}%
  \BibitemOpen
  \bibfield  {author} {\bibinfo {author} {\bibfnamefont {H.}~\bibnamefont
  {Wu}}, \bibinfo {author} {\bibfnamefont {C.~H.}\ \bibnamefont {Wan}},
  \bibinfo {author} {\bibfnamefont {Z.~H.}\ \bibnamefont {Yuan}}, \bibinfo
  {author} {\bibfnamefont {X.}~\bibnamefont {Zhang}}, \bibinfo {author}
  {\bibfnamefont {J.}~\bibnamefont {Jiang}}, \bibinfo {author} {\bibfnamefont
  {Q.~T.}\ \bibnamefont {Zhang}}, \bibinfo {author} {\bibfnamefont {Z.~C.}\
  \bibnamefont {Wen}}, \ and\ \bibinfo {author} {\bibfnamefont {X.~F.}\
  \bibnamefont {Han}},\ }\href@noop {} {\bibfield  {journal} {\bibinfo
  {journal} {Phys. Rev. B}\ }\textbf {\bibinfo {volume} {92}},\ \bibinfo
  {pages} {054404} (\bibinfo {year} {2015})}\BibitemShut {NoStop}%
\bibitem [{\citenamefont {Du}\ \emph {et~al.}(2014)\citenamefont {Du},
  \citenamefont {Wang}, \citenamefont {Yang},\ and\ \citenamefont
  {Hammel}}]{du2014systematic}%
  \BibitemOpen
  \bibfield  {author} {\bibinfo {author} {\bibfnamefont {C.~H.}\ \bibnamefont
  {Du}}, \bibinfo {author} {\bibfnamefont {H.~L.}\ \bibnamefont {Wang}},
  \bibinfo {author} {\bibfnamefont {F.~Y.}\ \bibnamefont {Yang}}, \ and\
  \bibinfo {author} {\bibfnamefont {P.~C.}\ \bibnamefont {Hammel}},\
  }\href@noop {} {\bibfield  {journal} {\bibinfo  {journal} {Phys. Rev. B}\
  }\textbf {\bibinfo {volume} {90}},\ \bibinfo {pages} {140407} (\bibinfo
  {year} {2014})}\BibitemShut {NoStop}%
\bibitem [{\citenamefont {Morota}\ \emph {et~al.}(2011)\citenamefont {Morota},
  \citenamefont {Niimi}, \citenamefont {Ohnishi}, \citenamefont {Wei},
  \citenamefont {Tanaka}, \citenamefont {Kontani}, \citenamefont {Kimura},\
  and\ \citenamefont {Otani}}]{morota2011indication}%
  \BibitemOpen
  \bibfield  {author} {\bibinfo {author} {\bibfnamefont {M.}~\bibnamefont
  {Morota}}, \bibinfo {author} {\bibfnamefont {Y.}~\bibnamefont {Niimi}},
  \bibinfo {author} {\bibfnamefont {K.}~\bibnamefont {Ohnishi}}, \bibinfo
  {author} {\bibfnamefont {D.}~\bibnamefont {Wei}}, \bibinfo {author}
  {\bibfnamefont {T.}~\bibnamefont {Tanaka}}, \bibinfo {author} {\bibfnamefont
  {H.}~\bibnamefont {Kontani}}, \bibinfo {author} {\bibfnamefont
  {T.}~\bibnamefont {Kimura}}, \ and\ \bibinfo {author} {\bibfnamefont
  {Y.}~\bibnamefont {Otani}},\ }\href@noop {} {\bibfield  {journal} {\bibinfo
  {journal} {Phys. Rev. B}\ }\textbf {\bibinfo {volume} {83}},\ \bibinfo
  {pages} {174405} (\bibinfo {year} {2011})}\BibitemShut {NoStop}%
\bibitem [{\citenamefont {Tian}\ \emph {et~al.}(2016)\citenamefont {Tian},
  \citenamefont {Li}, \citenamefont {Qu}, \citenamefont {Huang}, \citenamefont
  {Jin},\ and\ \citenamefont {Chien}}]{tian2016manipulation}%
  \BibitemOpen
  \bibfield  {author} {\bibinfo {author} {\bibfnamefont {D.}~\bibnamefont
  {Tian}}, \bibinfo {author} {\bibfnamefont {Y.~F.}\ \bibnamefont {Li}},
  \bibinfo {author} {\bibfnamefont {D.}~\bibnamefont {Qu}}, \bibinfo {author}
  {\bibfnamefont {S.~Y.}\ \bibnamefont {Huang}}, \bibinfo {author}
  {\bibfnamefont {X.~F.}\ \bibnamefont {Jin}}, \ and\ \bibinfo {author}
  {\bibfnamefont {C.~L.}\ \bibnamefont {Chien}},\ }\href@noop {} {\bibfield
  {journal} {\bibinfo  {journal} {Phys. Rev. B}\ }\textbf {\bibinfo {volume}
  {94}},\ \bibinfo {pages} {020403} (\bibinfo {year} {2016})}\BibitemShut
  {NoStop}%
\end{thebibliography}
\end{document}